\documentclass[
 reprint,
 amsmath,amssymb,
 aps,prl,eps,superscriptaddress,twocolumn,10pt
]{revtex4}

\usepackage{graphicx}% Include figure files
\usepackage{dcolumn}% Align table columns on decimal point
\usepackage{bm}% bold math
\usepackage{xcolor}
\begin{document}
\newcommand{\ajm}[1]{\textcolor{black}{\textit{\textbf{Andrew:} #1}}}
\newcommand{\why}{\textcolor{black}{\textit{\textbf{Andrew: Why?}}}}
\newcommand{\wdtm}{\textcolor{black}{\textit{\textbf{Andrew: What does this mean -- it is not clear?}}}}
\newcommand{\hb}[1]{\textcolor{black}{#1}}
\newcommand{\etal}{\textit{et al.}}
\newcommand{\hbcomm}[1]{\textcolor{black}{#1}}
\preprint{APS/123-QED}

\title{The importance of electronic correlations in exploring the exotic phase diagram of layered Li$_x$MnO$_2$}

\author{Hrishit Banerjee}
\email{hb595@cam.ac.uk}
\affiliation{Yusuf Hamied Department of Chemistry, University of Cambridge, Lensfield Road, CB2 1EW, Cambridge, UK.}
\affiliation{School of Metallurgy and Materials, University of Birmingham,Elms Road,  B15 2TT, Edgbaston, Birmingham, UK.}
\affiliation{The Faraday Institution, Quad One, Becquerel Avenue, Harwell Campus, Didcot, OX11 0RA, UK}

 %\email{}
% \altaffiliation[Also at ]{Physics Department, XYZ University.}%Lines break automatically or can be forced with \\
\author{Clare P. Grey}%
\affiliation{Yusuf Hamied Department of Chemistry, University of Cambridge, Lensfield Road, CB2 1EW, Cambridge, UK.}
\affiliation{The Faraday Institution, Quad One, Becquerel Avenue, Harwell Campus, Didcot, OX11 0RA, UK}
\author{Andrew J. Morris}
\email{a.j.morris.1@bham.ac.uk}
\affiliation{School of Metallurgy and Materials, University of Birmingham,Elms Road,  B15 2TT, Edgbaston, Birmingham, UK.}
\affiliation{The Faraday Institution, Quad One, Becquerel Avenue, Harwell Campus, Didcot, OX11 0RA, UK}

\date{\today}% It is always \today, today,
             %  but any date may be explicitly specified

\begin{abstract}
Using \textit{ab initio} dynamical mean-field theory we explore the electronic and magnetic states of layered Li$_x$MnO$_2$ as a function of $x$, the state of charge. Constructing real-space Wannier projections of Kohn-Sham orbitals based on the low-energy subspace of Mn $3d$ states and solving a multi-impurity problem, our approach focuses on local correlations at Mn sites.
The antiferromagnetic insulating state in LiMnO$_2$ has a moderate N\'{e}el temperature of $T_N=296\,K$ in agreement with experimental studies. 
Upon delithiation the system proceeds through a number of states:  ferrimagnetic correlated metals at $x$=0.92, 0.83; multiple charge disproportionated ferromagnetic correlated metals with large quasiparticle weights at $x$=0.67, 0.50, 0.33; ferromagnetic metals with small quasiparticle weights at $x$=0.17, 0.08
and an antiferromagnetic insulator for the fully delithiated state, $x=0.0$. 
At moderate states of charge, $x=0.67-0.33$,  a mix of +3/+4 formal oxidation states of Mn is observed, while the overall nominal oxidation of Mn state changes from +3 in LiMnO$_2$ to +4 in MnO$_2$. 
In all these cases the high-spin state emerges as the most likely state in our calculations considering the full $d$~manifold of Mn based on the proximity of $e_g$ levels in energy to $t_{2g}$. 
The quasiparticle peaks in the correlated metallic states were attributed to polaronic states, based on previous literature for similar isoelectronic JT driven materials,  \textcolor{black}{arising due to non-Fermi liquid type behaviour of the strongly correlated system}. 
\end{abstract}

\maketitle

%\tableofcontents

\paragraph{Introduction}
LiMnO$_2$ (LMO) \cite{Armstrong1996, CAPITAINE1996197, Vitins_1997} \hbcomm{has the potential to be} a low-cost, low-toxicity, high safety, and environmentally friendly alternative to the most popular rechargeable lithium-ion battery cathode material, LiCoO$_2$\cite{THACKERAY1983461, THACKERAY1984179, hackney, Armstrong1996, Freire2016, Chakraborty2020}, \hbcomm{however at 50\,\% delithiation LMO irreversibly transforms to a spinel (Fd$\bar{3}$m) phase, which causes significant reduction in the capacity and operating voltage, limiting its large-scale application \cite{Seymour2015, Seymour2016, Zhang_2020}. }

\textcolor{black}{The thermodynamically stable phase of LMO at ambient conditions is orthorhombic (Pmmn) \cite{GREEDAN1997} although a rhombohedral layered O3 structure, may be produced by ion exchange from NaMnO$_2$\cite{Armstrong1996,CAPITAINE1996197}.}
%it was shown independently by Armstrong et al \cite{Armstrong1996} and Capitaine et al \cite{CAPITAINE1996197} that a metastable phase of LMO, with a 
%
%LMO and LCO both adopt the same rhombohedral layered O3 stacking phase. 
%
The rhombohedral O3 structures comprise 
%
%In these structures, Co may be partially or completely replaced by combinations of different transition metals to give rise to 
%
a family of materials, including LiCoO$_2$ (LCO),  LiNiO$_2$ (LNO) \cite{Ohzuku_1993, Dahn_1991, BROUSSELY1995109} and  LiNi$_{1-x-y}$Mn$_y$Co$_x$O$_2$ (NMC) \cite{Makimura, YABUUCHI2003171, Xu2021, grey2021, Julien2016, STEPHAN2020, Chakraborty2020}. %where  $0 \leq x,y,z \leq 1$ and $x+y+z=1$. 
%Within the family of layered O3 materials, LMO has been suggested as an alternative to LCO, because of the low cost and low toxicity of Mn, as well as the high safety of LMO cathodes \cite{Chakraborty2020}.
%
Within this family LMO adopts the same oxygen stacking as LCO, but with symmetry reduced from rhombohedral (R$\bar{3}$m) to monoclinic (C2/m), as a result of a cooperative Jahn-Teller(JT) distortion. 
Based on magnetization data \cite{Tabuchi_1998, GREEDAN1997} pristine monclinic LMO is an antiferromagnetic insulator in its high spin state with relatively moderate Ne\'{e}l temperatures ($T_N\sim250\,K$)  due to the stabilisation of antiferromagnetism by the cooperative JT distortions~\cite{Tabuchi_1998}.

%The most practical and promising Li-ion cathode materials today are layered oxide materials, and in particular LiNi$_{z}$Mn$_y$Co$_x$O$_2$ (NMC) where  \hb{$0 \leq x,y,z \leq 1$ and $x+y+z=1$}\cite{Julien2016, STEPHAN2020, Chakraborty2020}.  A promising approach to discover the next generation electrodes \hb{for which it is imperative to have a balance of high voltage, good kinetics and high safety}, is to vary the composition of the battery electrode materials thereby optimising the desired parameters \cite{Xu2021, grey2021, Chakraborty2020}. The corner phases of ternary phase diagram generated from LiNiO$_2$ (LNO),  LiCoO$_2$ (LCO), and LiMnO$_2$ (LMO) form the 3 parent compounds of modern NMC cathodes.  Each corner phase has its own importance in terms of high voltage (LNO), better kinetics (LCO), and high safety (LMO) \ajm{cite}.  These layered oxides themselves being strongly correlated open $d$ shell materials have a predisposition  towards having a very intriguing correlated electronic phase diagram showing exciting phase transitions \hb{\cite{held06, held07, held10, Park14, Janson18, Lechermann19} due to the interplay of correlation with spin, orbital, and charge degrees of freedom}.

%\cite{Franchini2012, Kunes2017}.

Structural phase transitions during cycling have prevented the widespread adoption of \hbcomm{layered monoclinic} LMO as a cathode material.
Upon \textcolor{black}{50\% delithiation, layered phase to spinel structural phase transformation of LMO involves the migration of Mn ions, whilst the close-packed O lattice remains intact.}
Density-functional theory (DFT) calculations, combined with a hybrid eigenvector-following method to uncover the pathways, show that for the case of $x=0.5$, particular orderings of Li$^+$/Li$^+$ vacancy and Mn$^{3+}$/Mn$^{4+}$ ions play a significant role in this structural transformation \cite{Armstrong1996, Shao_Horn_1999, Reed_2001, Armstrong2004,Kim2012, Seymour2015}. 
Moreover the ionic pathways that give rise to the ordering and transformations are highly dependent on the inclusion of a Hubbard U term in the Kohn-Sham Hamiltonian, hinting that these structural changes may have a strong underlying electronic origin.
However, the role of electron correlations in the origin of the mixed-valence charge-ordered state, as well as the possibility of existence of such states at other states-of-charge remains unknown.

Electronic phase transitions occurring during cathode cycling\cite{Reimers_1992, Ohzuku_1994,Amatucci_1996, Delmas1999, Berg2021, Merryweather2021} significantly influence %\ajm{be more specific -- what ``influence'' does it have?} 
the reversibility of the (de)lithiation process, in terms of rate limiting formation and propagation of phase boundaries, lattice
mismatch and volume changes, all of which
lead to slow (de)lithiation kinetics hence affecting rate capability and stability, causing degradation of the active material \cite{Radin2017}.
%
%of all these cathode materials driven by electrochemical processes have been well studied \cite{Reimers_1992, Ohzuku_1994,Amatucci_1996, Delmas1999, Berg2021, Merryweather2021}, for they are believed to  
%
%As an example experimentally 
For example, a first-order metal-insulator transition, normally driven through doping \cite{werner07, Liebsch08} or pressure \cite{cryst8010038, Oliveira2021}, is driven electronically in Li$_x$CoO$_2$  \textcolor{black}{as a function of states-of-charge} $x$ \cite{Delmas1999, Merryweather2021, Berg2021}. 
%
%It is intriguing to note that the metal-insulator transition is driven here electrochemically as function of state-of-charge instead of the usual doping or pressure routes. 
%
Naturally the question arises whether such a electrochemically driven metal-insulator transition, or other exotic phases may also be observed in the analogous cathode material LMO. 
 
% \ajm{what is the subject of this paragraph, it is not clear from the first sentence. It needs redoing, I'm getting:
 %\begin{itemize}
  %   \item First sentence, MITs, Motts or MB methods?
  %   \item Second sentence, DFT and Motts (ok so we're talking about MTs?)
  %   \item Third sentence, DFT+DMFT, expolre a Mott -- (so what?)
  %   \item Fourth sentnece, Very exotic phase diagram.
  %   \item Fifth sentence, need MB methods for correlations.
 %\end{itemize}
 %}
 Electrochemical and magnetic phase transition %\ajm{find better word} 
 studies of complex layered oxides are made all the more robust through first principles modelling of the electronic and magnetic states of these materials. 
 Whilst the first order metal-insulator transition observed in LCO has been postulated to be a Mott transition by some DFT studies \cite{Marianetti2004, Berg2021}, a true Mott transition is difficult to capture within non-interacting DFT and warrants the use of many body methods like dynamical mean-field theory (DMFT) \cite{georges92, georges96, vollhardt, georges04, kotliar06, held-rev}.  
 %
 %Although  \cite{Marianetti2004, Berg2021} have proposed this to be a  Mott transition, it is well known \cite{georges92, georges96, vollhardt, georges04, kotliar06, held-rev}
 %\ajm{never say "well known" unless you have $>$3 papers cited here}  that a true Mott transition is difficult to capture within non-interacting Kohn-Sham DFT.% even if supplemented by an \textit{ad hoc} energy cost through a static ``Hubbard U'' term in the Kohn-Sham Hamiltonian. 
 %
 Recently a DFT + DMFT study to account for both static and dynamic correlations has been carried out to explore this Mott transition and associated compositional phase diagram in Li$_x$CoO$_2$ \cite{Marianetti2020}. 
 Here local interactions, which DMFT describes, are seen to
 strongly impact self energies, occupancies of $d$ orbitals, \textcolor{black}{phase stability} and electronic behaviour of LCO at various $x$. Pristine LCO is an insulator while delithiated phases are moderately correlated Fermi liquids with modest quasiparticle weights.
 %to describe this class of materials. 
 Quasiparticles like polarons are known to affect charge conductivity in batteries \cite{Sio1, Sio2}.
 %
% DFT+DMFT also correctly captures the phase stability of the system as a function of $x$, without a strong tendency for charge disproportionation at $x=0.5$ as seen in experiments and unlike what is found in DFT+U.
 %
 %A similar DFT+DMFT study predicts Co to be in its high-spin state \cite{Shorikov2016} although this has never been seen experimentally. \ajm{this is supposed to be a paragraph about why e need to model and why DMFT is the right method. Saying something has never been seen experimentally isnt a good argyment here -- can you rework, or find a better example?} %Thus these layered oxides being strongly correlated open $d$ shell materials have a predisposition towards having a very intriguing correlated electronic phase diagram showing exciting phase transitions due to the interplay of correlation with spin, orbital, and charge degrees of freedom. 
 %\ajm{I dont know what the paragraph is about}
 %``It is also to be noted''\ajm{fluff} here that 
 %LMO being isoelectronic to manganites and having similar Jahn-Teller distortions and possibly similar strong correlation effects due to open 3$d$ shell in $d^4$ already has a propensity towards a very exotic phase diagram like other manganites, where similar Mott transitions with emergence of strong quasiparticle peaks and polaronic behaviour and other novel phases have been captured by DFT+DMFT studies \cite{held06, held07, held10, banerjee-mplb, banerjee2020, banerjee2019, Park14, Janson18, Lechermann19}.
  %
 It is likely that similar treatment of strong correlations is essential for modelling plausible phase transitions in LMO.

 \textcolor{black}{One of the main bottlenecks to the use of DMFT in battery cathodes is that they need to be studied at different states-of-charge. This means dealing with multiple inequivalent sites, and hence solving multi-impurity problems. This has the disadvantage of having to solve very large density matrices which are not only very expensive, but may often have large off diagonal terms, leading to fermionic sign problems.}

In this letter we predict using DFT+DMFT the phase transitions in \hbcomm{layered monoclinic} Li$_x$MnO$_2$, as a function of $x$, and explore the temperature versus $x$ phase diagram. \textcolor{black}{We develop a computationally fast method to deal with multi-impurity DMFT calculations.}
%A deeper understanding of the phase transitions thereof provide keys to further improve the performance metrics of LMO 
%\ajm{don't like this -- could? we're reporting what we've done, it either DOES or it DOESN'T}. 
The electronic and magnetic state of pristine LMO are examined and compared to experimental results at temperatures below and above $T_N$. 
The system is then delithiated systematically and for each $x$ we present the spectral functions, magnetic properties, and transition temperatures. 
%
%The phase-diagram of $T$ versus $x$ is constructed based on this. 
%
We find  a metal-insulator Mott transition as a function of delithiation and the emergence of exotic states like ferrimagnetic correlated metals with large quasiparticle weights and charge-ordered ferromagnetic correlated metals with large quasiparticle peaks \textcolor{black}{arising due to non-Fermi liquid behaviour shown by polarons}. 
We clarify that origin of the charge disproportionated mixed valence states is due to different quasiparticle weights at different sites and postulate this ordering to be present for a range of $x$.  
\textcolor{black}{The pathways leading to structural transformations to spinel structure could be traced back to such charge disproportionated state.}

\paragraph{Computational Methodology}
\textcolor{black}{We develop a method to carry out multi-impurity DMFT calculations in a novel way by mapping  multi-impurity problems to equivalent number of single-impurity problems interacting through bath hybridisation by means of Wannier projections.}
\textcolor{black}{ Delithiating the system results in formation of crystallographically inequivalent Mn sites in the LMO supercell, which requires solving multi-impurity problems, involving solution of extremely large density matrices, which is computationally expensive, and has fermionic sign problem at low temperatures.
To make the problem tractable, Maximally Localised Wannier Functions (MLWF) were constructed for all structurally inequivalent sites, however each inequivalent site was solved as a single-impurity problem interacting through the bath with the other inequivalent sites by virtue of the constructed MWLF, and the bath hybridisation. This method works well for strongly localised systems and makes accurate calculations at low temperatures converge faster with much less noise, avoiding the fermionic sign problem.
Further details on parameters and comparison to conventional DMFT cf. SI}

\paragraph{Results}

\begin{figure}
    \centering
    \includegraphics[width=\columnwidth]{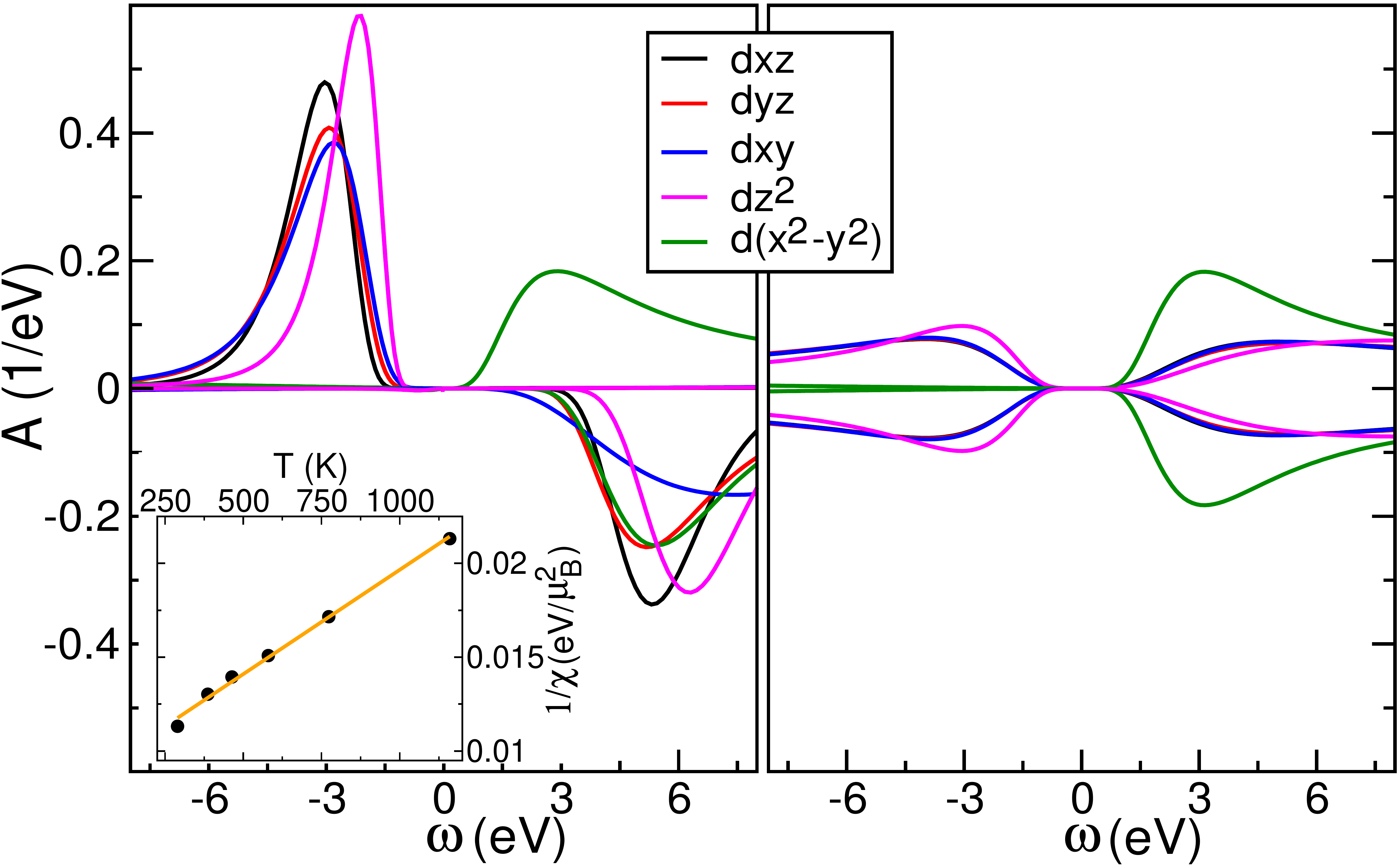}
    \caption{The DMFT spectral functions and magnetic calculations in LMO. Figure shows the DMFT spectral functions for both single-site spin-polarised (at T=58\,K, left panel) and paramagnetic (at T=580\,K, right panel) calculations. Inset in left panel shows the plot of 1/$\chi$ vs temperature on application of small magnetic fields at various temperatures. The straight line Curie-Weiss fit $\chi$=C/(T-$\theta$) for the data is shown. %\ajm{to save space could (b) be smaller?}
    }
    \label{fig2-new}
\end{figure}

To correctly account for dynamic correlation effects on Mn $d$ states, we first carry out single-impurity DMFT calculations for Mn $d$ based low-energy Hamiltonians defined in the basis of DFT-derived Wannier functions (cf. SI).

%\ajm{What is this paragraph about? XPS lineshapes, band gaps, states, impurity charges...?}
%
The paramagnetic state of pristine \hbcomm{layered monoclinic} LMO at T=580K shown by the paramagnetic spectral function %\ajm{of which structure?} 
in Fig.\,\ref{fig2-new} (a) (right) shows good agreement with XPS spectral line-shapes in experimental studies \cite{xiaohui}. 
Pristine LMO is an insulator \cite{Tabuchi_1998, Galakhov2000281} however experimental value of the electronic band gap is unknown. 
A wide range of band gap values from 0.3\,eV to 1.8\,eV have been reported from \textit{ab initio} calculations based on various choices of static Hubbard U in DFT+U calculations%\ajm{you mean based on various choices of U?} 
\cite{Galakhov2000281, Singh, NATHSHUKLA20061731, Ceder, cryst10060526}. 
We report a paramagnetic band gap of $\sim$0.6eV, with $U'=5eV$.
We find 3 slightly degeneracy broken $t_{2g}$ states that are half-filled, 1 $e_g$  state is half-filled and 1 $e_g$ state completely empty.
The impurity charge on Mn is 3.99 $|e|$ 
%\ajm{there is an ambiguity in using $e$, I prefer always to state charges in  which is then unambiguously signed} 
indicating Mn is in a $d^4$ (+3) formal oxidation state, again in agreement with XPS experiments \cite{xiaohui}.

\begin{figure}[th!]
    \centering
    \includegraphics[width=\columnwidth]{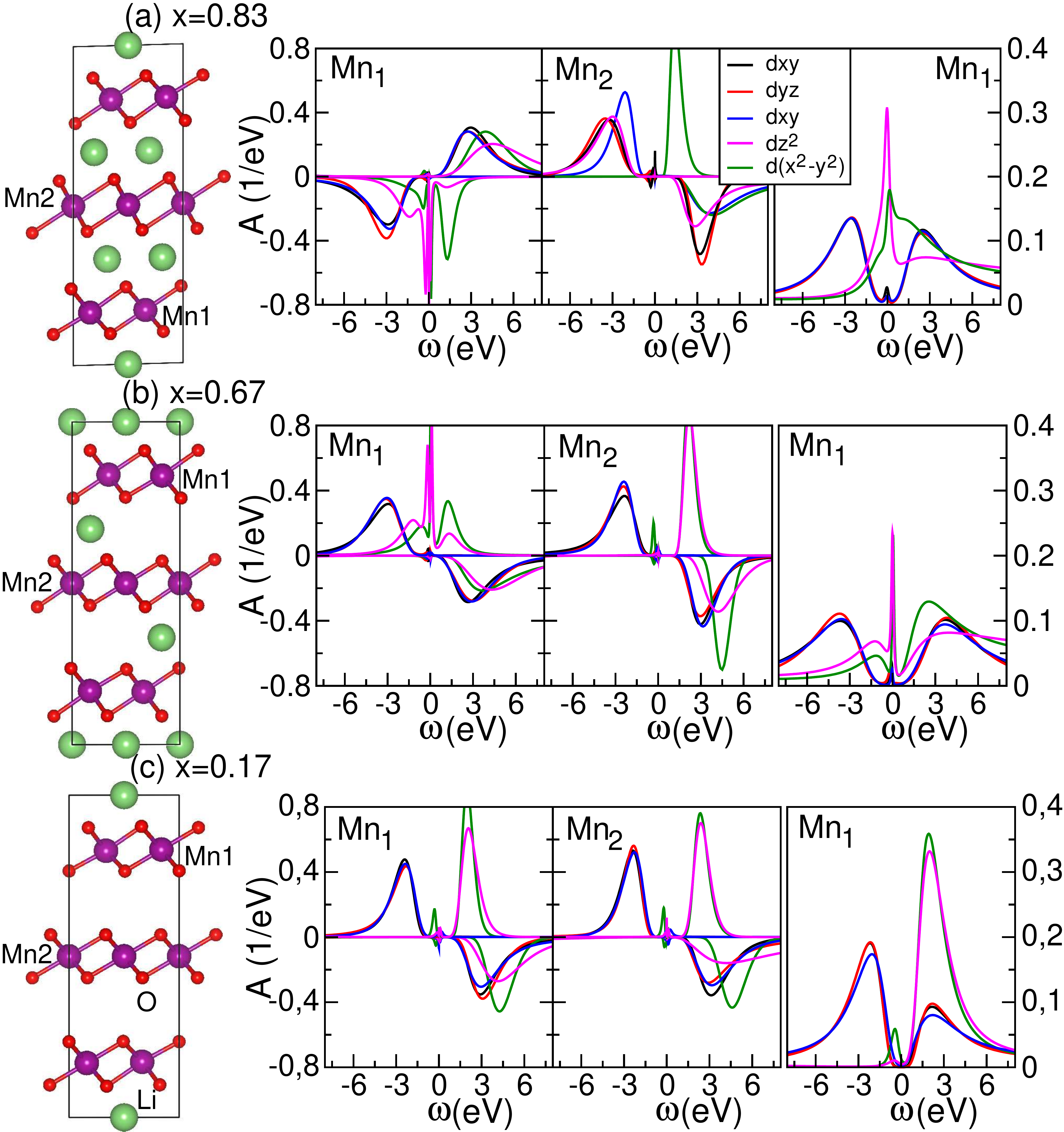}
    \caption{The structures and orbital resolved DMFT spectral functions for various states of charge showing the various phases as a function of $x$. The orbital spectral functions are denoted by the same colours as in Figure 1. 
    (a) shows the case of $x=0.83$.  
    %A large quasiparticle peak is seen at the chemical potential in Mn$_1$, being oppositely oriented to Mn$_2$, indicating a ferrimagnetic correlated metallic state, due to different multiplicities. 
    The top far right panel shows the paramagnetic spectral function for Mn$_1$.
    (b) shows the case of $x= 0.67$.  
    %All Mn sites have same orientation in spectral function. A large quasiparticle peak is seen at the chemical potential in Mn$_1$, indicating a strongly correlated metallic state. A mix of +3 and +4 states can be seen. 
    The middle far right panel shows the paramagnetic spectral function for Mn$_1$.
    (c) shows the case for $x=0.17$. 
    % Qualitatively all Mn have similar electronic structures and same orientation at all sites. A small quasiparticle peak is seen at Mn sites, with minor differences between sites 1 and 2. A majority of +4 oxidation states are seen. 
    The bottom far right panel shows the paramagnetic spectral function ofr Mn$_1$.
    %with a small quasiparticle peak. 
    }
    \label{fig3}
\end{figure}

Next, we proceed to explore the magnetism within single-impurity DMFT. 
For this purpose, we start from the paramagnetic solutions, add a symmetry breaking term in form of a spin-splitting term 
%\ajm{need a noun here?} 
in the real part of the self energies, and let the DMFT iterative cycle converge to a possible symmetry-broken solution with net ordered magnetic moment. 
We carry out the calculations at various different values of inverse temperature with
$\beta$ between 20 and 300\,eV$^{-1}$, where \hb{$\beta=\frac{1}{k_BT}$, where $k_B$ is the Boltzmann's constant}. 
At $\beta=20$\,eV$^{-1}$ ($T=580\,K$), the calculations are found to converge to a paramagnetic state, while upon reducing temperature, a transition to a magnetic solution is found. 
In Fig.(SI) 2 , we show a plot of the ordered average Wannier moments of Mn Wannier functions with the number of DMFT iterations. 
The ordered moments (saturating around 3.25\,$\mu_B$) are not stable, they %\ajm{"but rather" or "rather oscillaite"? I don't know what you mean here} 
oscillate as a function of iteration. 
This shows the propensity of the system towards antiferromagnetic fluctuations being present in the system \cite{bro}. The spectral function for the magnetic state shows insulating behaviour. 
The ordered saturation moment and antiferromagnetic behaviour is in good agreement with susceptibility experiments \cite{Tabuchi_1998}. 
Using a $U'=5\,eV$ within DMFT the antiferromagnetic band gap is $\sim$1.2\,eV.

%The reason for these oscillations is that antiferromagnetism naturally gives rise to two distinct sublattices $A$ and $B$, with the symmetry for the local Green's functions $G_{A,\sigma}(i\omega_n)=G_{B,\bar\sigma}(i\omega_n)$.  Here, in our ferromagnetic \hb{single impurity} setup, we do not have a sublattice structure, which corresponds to $G_{A,\sigma}(i\omega_n)=G_{B,\sigma}(i\omega_n)$. Since the impurity hybridisation function for sublattice $A$ is calculated from the self-energy and Green's function on sublattice $B$, it is clear that this leads to oscillations, when the sublattice structure and above given symmetry is not explicitly taken into account. 

%\ajm{Fine comment -- is it clear that this should be the start of a paragraoh?}
%From fluctuating moments it is difficult to estimate the $T_N$ N\'{e}el temperature for an antiferromagnetic to paramagnetic transition.
In order to determine the N\'{e}el temperature ($T_N$) we carry out susceptibility calculations with an external field applied on the system. 
%\ajm{what is a *ferromagnetic* external field?}
%
We vary the applied field from $0.01-0.03$\,eV 
%\ajm{this feels like the wrong unit, shoulnd't there be a per Angrstrom or something?} 
in steps of 0.01\,eV, and for each value of temperature we obtain the inverse slope of the magnetisation vs applied field within the linear regime. \hbcomm{Magnetic fields are applied as a split in energy and the value of magnetic field is added to the DFT Hamiltonian. The contribution $-h_\mathrm{field} \times \Sigma$ is added to diagonal elements of the Hamiltonian.}
This gives the inverse of the uniform susceptibility $1/\chi$ vs temperature $T$, as shown in  Fig.\ref{fig2-new} (b). 
By fitting the data to Curie Weiss law $\chi$=C/(T-$\theta$), where $C$ is Curie constant, $\theta$ is Weiss temperature in the high temperature regime we find $\theta=769\,K$ in excellent agreement with \hbcomm{experimentally reported values on two different samples $\theta=790\,K ,\ 540\,K$, \cite{Tabuchi_1998} and a calculated value of $\theta=790\,K$ \cite{Ceder}}. \hbcomm{This has been attributed to the stabilisation of antiferromagnetic state due to the cooperative JT distortion \cite{Ceder}. The calculated  N\'{e}el temperature ($T_N$) $\sim$296K, is also in agreement with experiments \cite{Tabuchi_1998}}.
We use the same set of parameters throughout for benchmarking of $T_N$ with experimental data.% for the rest of our calculations of the relevant transition temperatures.

Next we look at delithiating the pristine LMO by making two supercells of dimensions $3\times1\times1$ and $3\times2\times1$. 
This results in 6 and 12 symmetry equivalent Mn sites in case of $3\times1\times1$ and  $3\times2\times1$ respectively.
One Li atom is removed sequentially at each stage and various resulting structures are relaxed within DFT using the PBESol exchange-correlation functional. 
For the sake of brevity we do not enumerate all the structural details here however a significant contraction of lattice parameters and lattice volume is seen on delithiation, as also observed in experiments, as well as a reduction in JT distortion in the system is observed. 
The various fractions of $x$ considered here are $x=0.92, 0.83, 0.67, 0.50, 0.33, 0.17, 0.08,$ and $0.00$.
However, $x=0.00$ is an extreme case: removing all the Li from the structure is not possible experimentally and makes the structure unstable.

\begin{figure}[h!]
    \centering
    \includegraphics[width=\columnwidth]{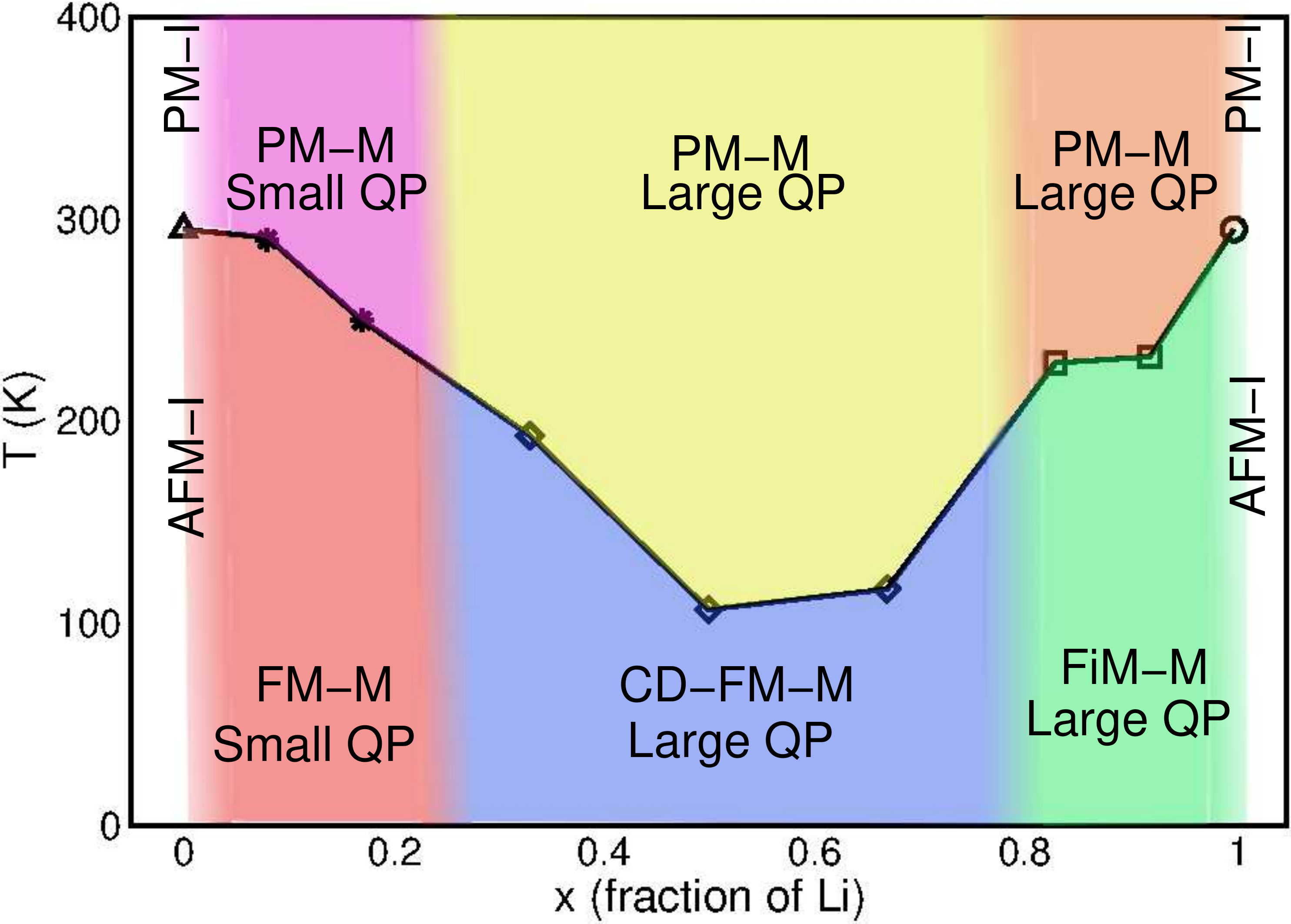}
    \caption{The different phases at different temperatures of Li$_x$MnO$_2$ as a function of Li content. Where AFM-I refers to antiferromagnetic insulator; FiM-M, ferrimagnetic metal; FM-M, ferromagnetic metal; PM-I, paramagnetic insulator; PM-M, paramagnetic metal; CD, charge disproportionated; and QP, quasiparticle. The different symbols on the line corresponds to the actual data points demarcating the different phases.}
    \label{fig4}
\end{figure}

% \ajm{we did this in the last paragraph. Moreover this isn't a way to start a paragraph -- We start WHAT?!}
In case of delithiation to $x=0.83$, the 6 symmetry equivalent Mn atoms in the pristine $3\times1\times1$ cell are now split by reduction of symmetry into have 4 structurally inequivalent types, Mn$_1$, Mn$_2$, Mn$_3$ and Mn$_4$ each with multiplicity 2, 1, 2 and 1 respectively. 
%
%Mn$_1$ and Mn$_3$ have 2 structurally equivalent sites each, and Mn$_2$ and Mn$_4$ has 1 structurally equivalent site each. 
%
%\ajm{Be clearer here, I think the subscript on Mn$_x$, i.e. the x is referring to the site in the pristine cell? If so, "the 4 symmetry equivalent Mn in the pristine cell are now split by the reduciton of symmetry into ..."  At the moment I can't quite understand what's being split into what, so I can'w write it for you}
%
We find a ferrimagnetic state in this case since from the DMFT spectral functions and from magnetic moments in Table 1 (SI) we see that Mn$_1$ with multiplicity=2 (Mn$_3$ spectral function is electronically similar to Mn$_1$) is oppositely oriented to Mn$_2$ with multiplicity=1 (Mn$_4$ spectral function is electronically similar Mn$_2$), as shown in Fig.\,\ref{fig3} (a).
%\ajm{other way round, "It is ferrimagnetic because... they're oppsosite orientation, as shown in Fig X. My comment still stands, you need to report the results in the paper -- NOT annotate the figures. This sentenece is the wrong way round.}
%
A large quasiparticle peak is seen at the chemical potential in Mn$_1$ and Mn$_3$ indicating a strongly correlated metallic state\cite{held-rev}. 
%\hbcomm{A strongly correlated metal which is in a non-Fermi liquid state is defined as a material with a quasiparticle peak at the chemical potential, where Coulomb repulsion U $\sim$ bandwidth W, unlike a free electron metal with a wide band at Fermi energy \cite{held-rev}.}
%
Thus these 4 structurally inequivalent types may thus be primarily be grouped into two categories - one with large quasiparticle and the other with small quasiparticle weights. 
This quasiparticle peak have a polaronic nature as suggested by previous studies on isoelectronic JT active manganites \cite{held, held07}. 
%
%Occupancy of different sites are are Mn$_1$=4.04, Mn$_2$=3.94, Mn$_3$=3.94, Mn$_4$=4.02 and averaged Wannier moments are at the different sites are Mn$_1$=-3.93, Mn$_2$=4.01, Mn$_3$=-3.75, Mn$_4$=3.95. \ajm{units?}
%
The spectral function for the $x=0.83$ filling is orbital resolved and shows that the $t_{2g}$ orbitals are partially filled and one of the $e_g$ orbitals is partially filled while the other $e_g$ orbital is empty, as shown in Fig \ref{fig3}(a). 
%\ajm{what figure are we looking at here? DON'T start the paragraph "in Fig X"! See above}
%
This implies an occupancy of $d^4$ and a formal oxidation state of +3. 
The occupancies and Wannier moments are given in Table 1 (SI).
The  paramagnetic phase at T=580 K corresponds to a strongly correlated paramagnetic metal with a large quasiparticle peak and non-Fermi liquid behaviour as seen from the paramagnetic spectral function shown in the top right panel of Fig \ref{fig3} (a).
The electronic structure for the case of $x=0.92$ is found to be very similar to $x=0.83$.
%, where we remove 1 Li from a $3\times2\times1$ supercell of LMO. Thus there are 11 Li out of 12 and a corresponding fraction of x=0.92. We have 6 structurally inequivalent sites in this case. The moments indicate a ferrimagnetic arrangement and a correlated metallic electronic state. We do not show the spectral functions for the sake of brevity.

Delithiated states with $x=0.67, 0.50, 0.33$ show similar electronic structure to each other across this range of $x$
%\ajm{to what? THIS IS STILL AMBIGUOUS. To each other, or to something else?} 
with certain slight differences, and belong to the same phase of charge-ordered ferromagnetic correlated metallic states. 
In case of $x=0.67$ and $x=0.33$, the 6 symmetry equivalent Mn atoms in the pristine $3\times1\times1$ cell are again split by reduction of symmetry into 4 structurally inequivalent types, Mn$_1$, Mn$_2$, Mn$_3$ and Mn$_4$, each with different multiplicities of 2, 1, 2, 1 respectively, however in case of $x=0.50$ there are 2 structurally inequivalent sites each with multiplicity 3. 
These sites may primarily be grouped again into two categories - one group with large quasiparticle and the other group with small quasiparticle weights. 
The $x=0.67$ state is ferromagnetic since all Mn sites have the same orientation in the spectral function, as shown in Fig.\,\ref{fig3} (b), and can also be seen from Moments in Table 1 (SI).
%
%This indicates a ferromagnetic arrangement. \ajm{backwards again}
%
A large quasiparticle peak is seen at the chemical potential in Mn$_1$ (multiplicity=2), Mn$_3$(multiplicity=2)  and Mn$_4$ (multiplicity=1).
Mn$_3$ and Mn$_4$ not shown here for the sake of brevity. 
The middle right panel of Fig. \ref{fig3} shows the paramagnetic spectral function as a strongly correlated metal for Mn$_1$, for the case of $x=0.67$.
Large quasiparticle weights are also seen on Mn$_1$, with multiplicity of 3 (for $x=0.50$) and on Mn$_1$ with multiplicity 2 for $x=0.33$, indicating a strongly correlated metallic state in all cases. 
%
%
%Occupancies in case of $x=0.67$ are Mn$_1$=4.01, Mn$_2$=3.98, Mn$_3$=3.83, Mn$_4$=3.07 and moments are Mn$_1$=3.91, Mn$_2$=3.87, Mn$_3$=3.65, Mn$_4$=2.99, for x=0.50 are, Mn$_1$=3.89, Mn$_2$=3.07, and moments Mn$_1$=3.75, Mn$_2$=2.99, and for x=0.33, occupancies are Mn$_1$=3.07, Mn$_2$=3.06, Mn$_3$=3.92, Mn$_4$=3.07, and moments Mn$_1$=2.97, Mn$_2$=2.98, Mn$_3$=3.77, Mn$_4$=2.99.
%\ajm{is this enough to ponder a table?}
%
From Table 1 (SI)  a mix of +3 and +4 states are seen in all these cases of $x=0.67, 0.50, \, \and 0.33$  %\ajm{all cases of what? You're referring to  $x=0.67, 0.50, 0.33$ or $\forall x$?} 
with different fractions of +3 and +4 Mn being present in the system. \hbcomm{It is also seen from the spectral functions that this mixed oxidation state and corresponding charge disproportionation originates due to different quasiparticle weights on different sites.}
 Spectral functions obtained for $x=0.50$, and $0.33$ are seen to be qualitatively similar to the case of $x=0.67$. 
 %\ajm{Simlar to what? Something else, or each other?}
%
% \ajm{whay are we talking about  $x=0.67$? We'd nmoved on to 0.5 and 0.33 -- please reorganise logically}
 
In case of the delithiated state with $x=0.17$ the 6 symmetry equivalent Mn atoms in the pristine $3\times1\times1$ cell are also split by reduction of symmetry into 4 structurally inequivalent types, Mn$_1$, Mn$_2$, Mn$_3$ and Mn$_4$ each with multiplicity 2, 1, 2 and 1 respectively. 
Qualitatively all Mn have similar electronic structures at all sites. 
A ferromagnetic arrangement is observed here since all Mn sites have same orientation  as can be seen from the spectral function in fig. \ref{fig3} (c) as well as in Table 1 (SI).
A small quasiparticle peak is seen at the chemical potential in each of Mn$_1$, Mn$_2$ and Mn$_3$ indicating a moderately correlated metallic state. 
%
%Occupancies of different Mn sites are Mn$_1$=3.06, Mn$_2$=3.05, Mn$_3$=3.07, Mn$_4$=3.05, and moments are Mn$_1$=2.99, Mn$_2$=2.98, Mn$_3$=2.98, Mn$_4$=2.98.
A majority of +4 oxidation states are seen. 
There is no charge disproportionation/ordering in this case. 
The bottom right panel \ref{fig3} (c) shows the paramagnetic weakly correlated metallic state with a small quasiparticle peak. 
Similar spectral functions and magnetic structures are obtained in case of $x=0.08$. 
%\ajm{I think my suggestions above apply equally to this paragraph too}

%Finally to explore a very high state of charge we return to the large $3\times2\times1$ supercell and remove 11 Li out of 12 and explore the high charge state of x=0.08. In this case there are 6 structurally inequivalent Mn atoms, each with multiplicity 2, 2, 1, 1, 4, and 2 respectively. When solved we obtain occupancies Mn$_1$=3.08, Mn$_2$= 3.08, Mn$_3$=3.06, Mn$_4$=3.06, Mn$_5$=3.04, Mn$_6$=3.04. and moments, Mn$_1$=2.98, Mn$_2$=2.97, Mn$_3$=2.96, Mn$_4$=2.97, Mn$_5$=2.98, Mn$_6$=2.98. This indicates a ferromagnetic arrangement again. The spectral function shows a correlated metal with very small quasiparticle peak.
%Fully delithiating the system to \hbcomm{$x=0.0$} results in an unstable structure of MnO$_2$ (cf. SI). 
%
%In our DMFT calculations we find this to be a system with antiferromagnetic fluctuations being present in the system (cf. SI). 

\paragraph{Discussion}
\textcolor{black}{The exotic phase diagram and associated phase transitions in Li$_x$MnO$_2$ as a function of $x$ vs temperature, is shown in Fig \ref{fig4}. The y axis shows the phase change from magnetic to paramagnetic states at each $x$  driven by competition of thermal fluctuations with magnetic order. The change of magnetic phases along $x$ is more nuanced.}
\textcolor{black}{
JT distorted materials show large superexchange which favours anti-ferromagnetism. This is the case of the pristine material $x=1$, which is a JT driven antiferromagnetic insulator. As the system is delithiated, JT distortion reduces on different Mn$O_6$ octahedra, which in turn indicates a reduction in monoclinicity. This reduction in JT favours ferromagnetism instead of anti-ferromagnetism. At $x=0.83$ there are few sites with reduced JT distortion and hence ferromagnetic exchange whereas majority of other sites still have significant JT distortion and hence show Mn-Mn antiferromagnetic superexchange, leading to an overall ferrimagnetic state. When further delithiated to $x=0.67$, and beyond, the overall reduction of JT distortion now favours FM double exchange, between different occupancies at majority of Mn sites. Reduction in JT distortion is usually associated with ferromagnetism particularly in Manganites \cite{banerjee2019, banerjee2020, spaldin}.}

\begin{figure}
    \centering
    \includegraphics[width=\columnwidth]{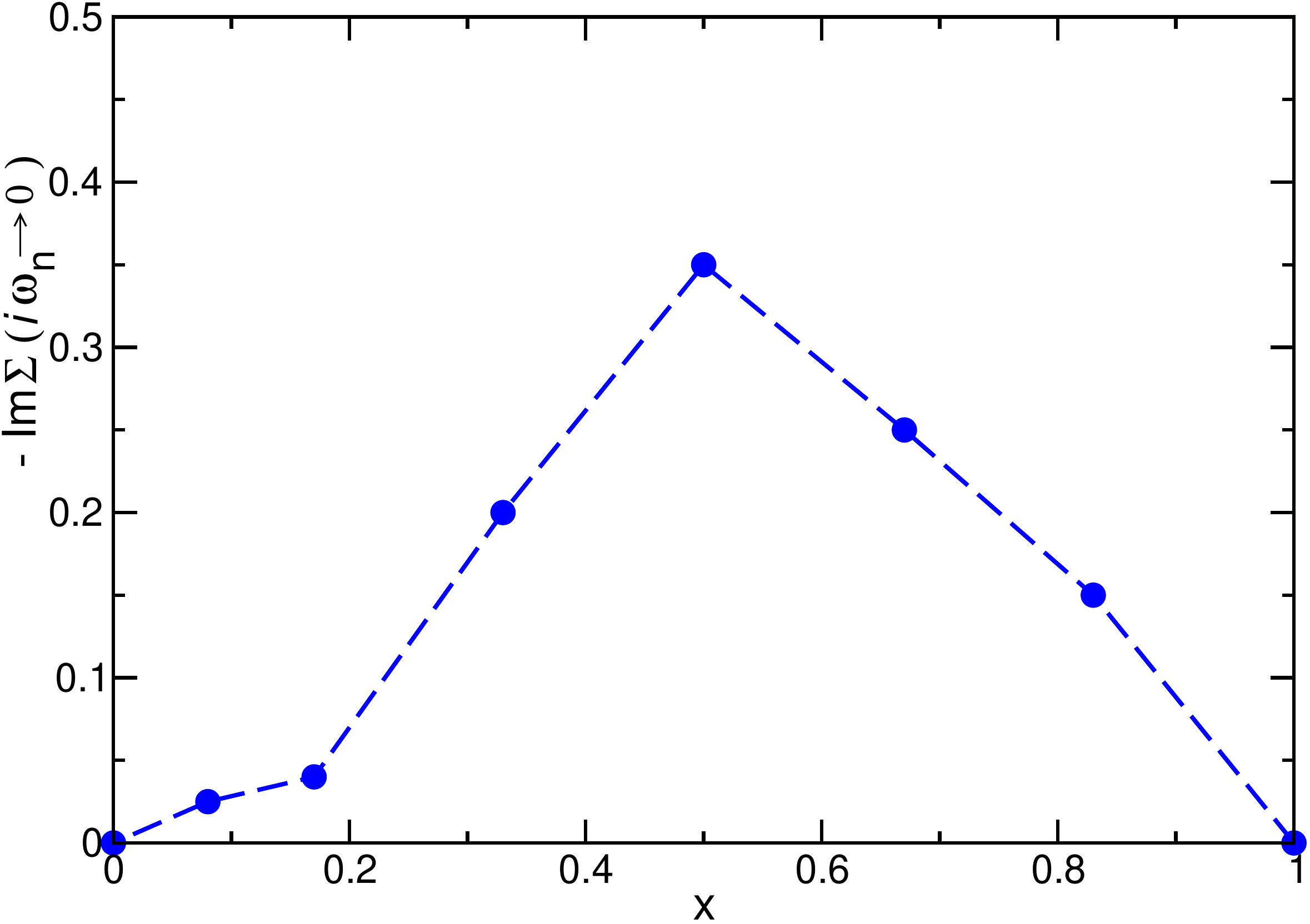}
    \caption{\textcolor{black}{Figure showing the plot of -Im $\Sigma$ ($\omega \rightarrow 0$) which gives a measure of coherence/incoherence behaviour as a function of state of charge plotted for the sites with largest quasiparticle weights.  A finite value of this quantity indicates a strongly correlated non-Fermi liquid type metallic behaviour with a large quasiparticle weight indicating the presence of large polarons. An infinitesimally small value tends to indicate a weakly correlated coherent Fermi liquid type behaviour.}}
    \label{fig5}
\end{figure}

\textcolor{black}{To elucidate the origin of polaronic quasiparticles on delithiating we look at the crossover from coherence (Fermi-liquid) to incoherence (non-Fermi liquid) behaviour as a function of $x$. We plot  $-Im \Sigma(\omega \rightarrow 0)$, $\Sigma$ is self energy, as a function of $x$ shown in Fig. \ref{fig5} which gives the measure of the coherence/incoherence behaviour of the quasiparticle states \cite{bro}. A finite value of this quantity indicates a strongly correlated incoherent state - such as a polaron in this case, which is indicative of non-Fermi liquid behaviour, in ground state, not captured by DFT. This manifests as a large localised peak in the spectral function obtained from interacting Green’s functions, which is known as a quasiparticle peak. Large quasiparticle polaronic states arise on delithiation on some Mn sites owing to non-Fermi liquid behaviour of the system. Charge disproportionation into Mn$^{3+}$/Mn$^{4+}$ states result from large quasiparticles on some sites and small quasiparticles on others.
 $-Im\Sigma(\omega \rightarrow 0)$ is shown for the site with largest quasiparticle weights seen in spectral functions. As we show an increase in incoherent or non-Fermi liquid type behaviour is seen with delithiation from $x$=0.83 to $x$=0.33, which indicates that these are strongly correlated quasiparticle states, and explains the origin of the polaronic states.
Doped manganites in general have been known to show polaronic quasiparticle peaks\cite{held07}. }

%

%System has an impurity charge=2.98 with 3 half filled $t_{2g}$ orbitals, 2 empty $e_g$ orbitals. The JT distortions vanish completely as shown in Fig S2. The average Wannier moments fluctuates with a maximum saturation value of $\sim 2.50\mu_B$. This indicates presence of antiferromagnetic fluctuations in the system. The system is insulating.

\paragraph{Conclusion}
In summary we have shown the exotic phase diagram and phase transitions in Li$_x$MnO$_2$ as a function of state-of-charge $x$, and temperature. \textcolor{black}{Our study shows a novel method of carrying out multi-impurity DMFT calculations in cathode materials in a computationally tractable way, thus opening up the field of battery physics in terms of strong correlations based many-body studies. It demonstrates an electrochemical method of tuning strongly correlated phase transitions and also shows emergence of polaronic quasiparticle states which contribute to charge disproportionation and which eventually lead to structural transformations seen in this material, and helps us in understanding degradation in cathodes.}

We find that an antiferromagnetic insulating state emerges in LiMnO$_2$, with a Weiss temperature of $\theta=769\,K$, in its $+3$ formal oxidation state, in high-spin configuration, in excellent agreement with experimental measurements. 
As the system is delithiated  at fractions of $x=0.92$, and $0.83$ we predict a ferrimagnetic correlated metallic state thus observing a metal-insulator transition, similar to that in LCO.  
From the DMFT spectral function this metal-insulator transition is seen to be of a Mott type transition with a large quasiparticle weight in the strongly correlated metallic state. 

At fractions of $x=0.67, 0.5$, and $0.33$ we find the system to be in ferromagnetic strongly correlated metallic state with a mix of +3/+4 formal oxidation states. 
\textcolor{black}{The pathways leading to the structural transformation to low energy spinel structure, originates with orderings of Li$^+$/Li$_{vac}$ and Mn$^{3+}$/ Mn$^{4+}$ mixed oxidation states at $x=0.50$ \cite{Seymour2015}.}
We postulate that this \hbcomm{Mn$^{3+}$/ Mn$^{4+}$ charge disproportionation} occurs due to the different quasiparticle weights at different sites.

At a high state of charge with $x=0.17$, we find a ferromagnetic correlated metal with a small quasiparticle peak near a formal oxidation state of +4. 
%
%Finally on fully delithiating to MnO$_2$ an antiferromagnetic insulating state is observed again, wherein the structure is albeit unstable. 

An overall nominal oxidation state change of Mn from +3 in LiMnO$_2$ to +4 in MnO$_2$ is observed, through stages of mixed oxidation states. 
In all these cases the high-spin state emerges as the most likely state considering the full $d$ manifold of Mn in the model. 

The quasiparticle peaks in the correlated metallic states were attributed to polaronic states, \textcolor{black}{owing to incoherent non-Fermi liquid behaviour,} and the correlated metallic systems may be thought of as polaronic hopping semiconductors. 

Our novel and faster method of carrying out multi-impurity DMFT calculations involves calculating Wannier projections for all impurities and solving each separately keeping others in the bath, interacting through the bath hybridisation. 

To the best of our knowledge a systematic examination of the electrochemically driven Mott transition and the full phase diagram of LMO from a correlations based perspective does not exist, nor does an explanation for the observed orbital ordering at $x=0.50$ and our study is expected to give rise to further experimental studies exploring these aspects of this very important cathode material through possible \textit{in-situ} XPS+BIS spectra and magnetic susceptibility measurements while charging and discharging of the LMO cathodes. 
%\textcolor{blue}{It is also to be noted that without many body methods like DMFT it is not possible within DFT or DFT+U methods to capture this rich quasiparticle based physics at moderate values of U $\approx$ W, which dominates the physics of battery cathodes on charging-discharging processes both at high (paramagnetic) and low (magnetically ordered) temperature regimes.}

    \begin{acknowledgments}
The authors acknowledge fruitful discussion with Annalena R. Genreith-Schriever.
This work has been funded by the Faraday Institution degradation project (FIRG001, FIRG024). 
Calculations have been performed on the odyssey cluster, the CSD3 cluster of the University of Cambridge.
Generous computing resources were provided by the Sulis HPC service (EP/T022108/1), and networking support by CCP‐NC (EP/T026642/1), CCP9 (EP/T026375/1), and UKCP (EP/P022561/1).
\end{acknowledgments}

\bibliography{main}

\end{document}